\documentclass[letterpaper,twocolumn,10pt]{article}
\usepackage{artifact/templates/usenix2019_v3}
\usepackage[T1]{fontenc}
\usepackage{graphicx}
\usepackage{booktabs}
\usepackage{amsmath}
\usepackage{microtype}
\usepackage{url}
\usepackage{pifont}
\usepackage{capt-of}
\hypersetup{hidelinks,pdftitle={Range: Decoupling KV-Cache Reuse Granularity from Storage Work}}
\title{Range: Decoupling KV-Cache Reuse Granularity from Storage Work}
\author{
{\rm Dingsen Shi$^{1}$, Jiahao Zhang$^{2}$, Liqiang Lv$^{2}$, Boming Huang$^{2}$, Yanbo Dai$^{1}$,}\\
{\rm Ying He$^{1}$, Donger Chen$^{1}$, Harry Fu$^{1}$, Wang Feng$^{1}$, Song Fu$^{1}$}\\[4pt]
$^{1}$University of North Texas \quad $^{2}$SageShield Co., Ltd.\\[2pt]
{\small\tt \{DingsenShi, YanboDai, YingHe, DongerChen, HarryFu, WangFeng\}@my.unt.edu, Song.Fu@unt.edu}\\
{\small\tt jiahaozhang.sci@gmail.com, lvliqiang@sageshield.com.cn, bominghuang92@gmail.com}
}

\begin{document}
\maketitle
\begin{abstract}
LLM serving reuses cached key-value (KV) state to skip prefill, and growing
contexts push that state past GPU HBM and CPU DRAM onto NVMe SSDs, placing
prefix restore on the critical path of time-to-first-token (TTFT). We find that
on a well-configured host-issued path the SSD is not the bottleneck: restoring
the same 6\,GiB of KV takes 2.1\,s as 192 objects but 3.0\,s as 6,144 objects,
and more under layerwise loading, because every layer of the storage
stack---lookup, allocation, I/O submission, GPU delivery, store, index---charges
per logical object while the bytes stay the same. This cost grows with the KV
footprint: longer contexts, deeper models under layerwise loading, and tensor
parallelism each multiply the object count. Prior systems move this per-object
work from the CPU to the GPU; the work remains per object.

This paper presents Range, a KV-cache storage engine for vLLM and LMCache
built on one principle: every cost on the storage path scales with bytes or
physical batches, never with logical objects. \emph{Range tickets} replace
per-key lookup and ownership with token metadata and physical leases;
\emph{byte stripes} make the on-media layout independent of reuse granularity;
\emph{completion-owned GPU transfer banks} batch delivery and extraction and
release each layer to attention as it lands. Range keeps the kernel cuFile path
and changes four lines of vLLM. On a fixed 6\,GiB restore, Range's median TTFT
stays at 1.79--1.80\,s across a $32\times$ range of object counts and under
layerwise delivery, within 70\,ms of the prefix already being resident in GPU
memory, while tuned in-process LMCache rises from 2.12\,s to 3.01\,s. Range
improves on the fastest multiprocess DRAM-bounced path by 0.44\,s at coarse
granularity while moving no payload through host memory, and holds layerwise
reuse at the cost of coarse-grained reuse, where the best stock path takes
2.24\,s.
\end{abstract}

\section{Introduction}
\label{sec:introduction}

Prefix caching has become a standard optimization in LLM serving: the
key-value (KV) state computed for a prompt prefix is kept and reused, so a
request that shares that prefix skips its prefill and pays only for its new
tokens~\cite{vllm-connector-contract,lmcache-gds-docs,cachegen,cacheblend}.
The benefit is bounded by how much KV a server can retain. As context windows
and concurrency grow, KV footprints exhaust GPU HBM, then CPU DRAM, and serving
stacks add NVMe SSDs as the next tier to keep hours of conversation and agent
state reusable~\cite{mooncake,lmcache-mp-source,tutti,strata,dualblade}. Once
KV lives on SSD, restoring a cached prefix sits on the critical path of time to
first token (TTFT): the request cannot start until its prefix is back in GPU
memory.

The received view is that SSD-backed KV reuse is slow because of the path
between the SSD and the GPU. Tutti attributes the slowdown to the fragmented,
page-based layout of GPU memory, which turns a restore into many small I/Os,
and to the CPU that must initiate each of them, and concludes that host-issued
I/O---even with GPUDirect Storage---is the bottleneck~\cite{tutti}. Its answer,
following BaM and GeminiFS, moves I/O issuance and completion onto the
GPU~\cite{bam,geminifs,tutti,copilotio}. LMCache's multiprocess tier takes the
opposite route: it keeps issuance on the host, moves it into a native worker
pool that prefetches at scheduling time, and stages every byte through a
pinned DRAM slab on its way to the GPU~\cite{lmcache-mp-source}.

\begin{figure}[t]
  \centering
  \includegraphics[width=\columnwidth]{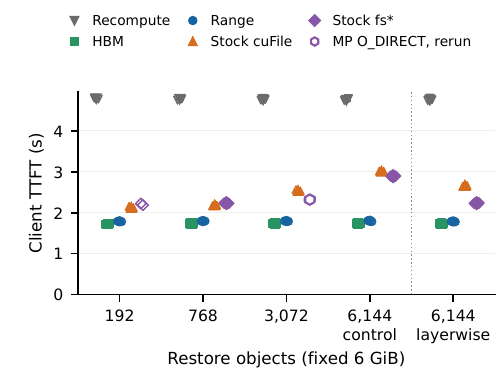}
  \caption{Client time to first token for the fixed-payload object-count
  sweep: the same cached prefix restores 6\,GiB, with logical granularity
  varied independently of layerwise delivery. Open marks are individual
  requests; filled marks and ranges show the median and observed minimum
  and maximum for complete cells. Stock fs* is the multiprocess filesystem
  path at 192--3,072 objects and O\_DIRECT at the finer points; missing
  multiprocess cells are fixture failures, and a separate rerun marker
  follows the repair. The HBM-resident control uses the same request.}
  \label{fig:primary-ttft}
\end{figure}

We find that on a well-configured host-issued path the SSD is not the
bottleneck, and neither is the processor that issues the I/O. Restoring the
same 6\,GiB prefix through tuned in-process LMCache over cuFile takes
2.124\,s when the prefix is cached as 192 objects and 3.012\,s as 6,144
objects; under real layerwise loading, which is how a serving engine creates
fine granularity, it takes 2.664\,s (Figure~\ref{fig:primary-ttft}). The bytes
are identical. What changes is the number of \emph{logical objects}, and every
layer of the storage stack---lookup, allocation, I/O submission, GPU delivery,
store submission, index maintenance---charges per object. This cost grows with
the KV footprint itself: longer contexts add objects, layerwise loading
multiplies them by the number of layers, and tensor parallelism multiplies
them again. A 128K-token prefix on an 80-layer model under layerwise loading
with eight-way tensor parallelism is over 300,000 objects for one request.
Because the stock cost is per object while the media cost is per byte, the
gap between them widens with the KV footprint: in our measurements, doubling
the payload at a fixed object size doubles the stock path's host-side time,
while the media time grows only with the bytes. Moving the per-object work to
the GPU, as GPU-issued designs do, leaves it per object; hiding it behind a
DRAM stage, as the multiprocess tier does, pays for it in host memory
bandwidth and pinned capacity and still has no layerwise path.

This paper presents Range, a KV-cache storage engine for vLLM and LMCache
built on one principle: \emph{every cost on the storage path scales with bytes
or physical batches, never with logical objects}. A logical object names a
reusable unit of the prefix; a physical byte batch moves its payload. Range
keeps the first for what it is good for---deciding where a partial hit
ends---and removes it from everything else. Three mechanisms carry that
representation end to end:

\begin{itemize}
\item \textbf{Range tickets} (\S\ref{sec:design}) replace per-key lookup and
      ownership with token metadata, physical-pack generations, and request
      leases. Restoring per-key indexing costs 23\,ms at 32\,MiB objects and
      911\,ms under layerwise loading.
\item \textbf{Physical byte stripes} store KV in fixed-size stripes inside
      1\,GiB packs, so the on-media layout and the number of native I/O calls
      are independent of reuse granularity: a 6\,GiB restore is 192 stripe
      reads at every object count, and a 2016\,MiB suffix store is 64 native
      writes.
\item \textbf{Completion-owned transfer banks} batch GPU delivery and direct
      GPU extraction through eight 32\,MiB banks that retain ownership until
      their consumers finish, and release each layer to attention as it
      lands. Replacing layer-ready delivery with a whole-prefix barrier costs
      roughly 250\,ms at every point.
\end{itemize}

Range is implemented as an LMCache storage backend and vLLM connector
(\S\ref{sec:implementation}). It keeps the kernel cuFile path, changes four
lines of vLLM, and leaves the model, the KV representation, and attention
untouched. We evaluate it on a fixed 6\,GiB restore across a $32\times$ range
of logical object counts and real layerwise delivery, against tuned
in-process cuFile, direct I/O, the multiprocess filesystem tier, recompute,
and a control in which the prefix is already resident in GPU memory
(\S\ref{sec:method}--\S\ref{sec:evaluation}). Range's median TTFT stays at
1.788--1.803\,s at every point, within 47--68\,ms of the resident-prefix
control, while stock in-process paths rise from 2.124\,s to 3.012\,s. At
coarse granularity it improves on the multiprocess filesystem path by
438\,ms while moving no payload through host memory; under layerwise loading
it is 447\,ms faster than the best stock path (direct I/O) and 876\,ms faster
than in-process cuFile.

This paper makes the following contributions:
\begin{itemize}
\item A diagnosis of SSD-backed KV reuse in which the cost is per logical
      object rather than per byte or per issuing processor---and therefore
      grows with context length, model depth under layerwise loading, and
      tensor parallelism---together with a fixed-payload sweep that isolates
      it.
\item Range tickets, physical byte stripes, and completion-owned transfer
      banks: a storage representation that keeps lookup, I/O, delivery,
      store, and lifetime management per byte or per batch.
\item An implementation in LMCache and vLLM, and an evaluation showing that
      fixed-payload KV reuse becomes insensitive to object count---so that
      Range's cost, unlike the stock path's, does not grow as finer
      granularity, deeper models, or larger contexts multiply the object
      count---with the cost of each mechanism attributed by ablation.
\end{itemize}

\section{Background and Motivation}
\label{sec:background}
\newcommand{\MotNativePayload}{6}
\newcommand{\MotNativeObject}{32}
\newcommand{\MotNativeRepetitions}{5}
\newcommand{\MotNativeOne}{15.61--16.10}
\newcommand{\MotNativeFour}{22.91--23.10}
\newcommand{\MotPrefixPayload}{6}
\newcommand{\MotStockCoarse}{2.713--2.774}
\newcommand{\MotStockFine}{3.738--3.810}
\newcommand{\MotStockLayerwise}{4.455--4.620}
\newcommand{\MotStoreBytes}{2016}
\newcommand{\MotStoreWrites}{64}
\newcommand{\MotStoreGathers}{8}
\newcommand{\MotStoreLeases}{8}
\newcommand{\MotStoreObservations}{25}
\newcommand{\MotValidationCalls}{32}
\newcommand{\MotValidationDeferredInside}{0}
\newcommand{\MotValidationPeak}{192}
\newcommand{\MotValidationTerminal}{0}
\newcommand{\MotStockCoarseObjects}{192}
\newcommand{\MotStockMediumObjects}{768}
\newcommand{\MotStockSmallObjects}{3072}
\newcommand{\MotStockFineObjects}{6144}
\newcommand{\MotStockLayerwiseObjects}{6144}
\newcommand{\MotCountCoarseObjects}{192}
\newcommand{\MotCountMediumObjects}{768}
\newcommand{\MotCountSmallObjects}{3072}
\newcommand{\MotCountFineObjects}{6144}
\newcommand{\MotCountLayerwiseObjects}{6144}
\newcommand{\MotNativeTwo}{22.65--22.97}
\newcommand{\MotCpuStockCurrent}{2.739--2.761}
\newcommand{\MotCpuStockDeferred}{2.110--2.155}
\newcommand{\MotCpuRangeCurrent}{2.432--2.551}
\newcommand{\MotCpuRangeDeferred}{1.772--1.787}
\newcommand{\MotReadyCoarseStock}{0.9844--1.0027}
\newcommand{\MotReadyCoarseRange}{0.9908--1.0111}
\newcommand{\MotReadyLayerwiseStock}{0.9695--1.0181}
\newcommand{\MotReadyLayerwiseRange}{0.9814--1.0195}

This section first reviews how LLM inference produces and reuses KV state and
how serving engines lay it out in GPU memory (\S\ref{sec:bg-llm}). It then
measures where an SSD-backed reuse path loses time and shows that the loss is
charged per logical object, not per byte (\S\ref{sec:bg-loss}). It surveys
how existing systems answer that loss and where each leaves the per-object
work in place (\S\ref{sec:bg-prior}), and closes with the three challenges a
storage path must meet to make its work independent of object count
(\S\ref{sec:bg-challenges}), which the next section's mechanisms address one
by one.

\subsection{LLM inference, KV cache, and paged layout}
\label{sec:bg-llm}

\paragraph{Prefill, decode, and the KV cache.}
Transformer inference has two phases. Prefill processes the whole prompt in
parallel and computes, for every layer, the key and value tensors of every
token; it is compute-bound and ends when the first token is emitted, so its
cost is time to first token (TTFT). Decode then generates one token per step,
attending over the keys and values of all previous tokens. Engines keep those
tensors in a KV cache so that decode does not recompute them. The cache is not
bound to one request: two prompts that share a prefix produce identical KV for
that prefix, so a serving engine that retains it can skip the shared part of
prefill entirely. This prefix caching is a standard optimization in vLLM,
SGLang, and LMCache, and it turns a compute-bound phase into a data-movement
problem---the cached prefix must be back in GPU memory before the new tokens
can be processed~\cite{vllm-connector-contract,lmcache-gds-docs,cachegen}.

\paragraph{Paged KV memory and its objects.}
Because sequences grow and vary in length, engines allocate KV in fixed-size
blocks rather than contiguous tensors. vLLM's paged layout stores each layer's
keys and values in blocks of a few tokens (16 in our configuration), indexed
through a block table per request~\cite{vllm-connector-contract}. A cached
prefix is therefore not one buffer but a set of blocks per layer, and a KV
tier outside the GPU has to name and move that set. LMCache groups tokens into
chunks (256 tokens by default) and names each chunk by a hash of its token
prefix; with layerwise loading the unit becomes one chunk of one layer, so a
32-layer model produces 32 objects per chunk. For Llama-3.1-8B, whose KV is
128\,KiB per token, a 256-token chunk is 32\,MiB and a per-layer object is
1\,MiB; a 49,152-token prefix is 6\,GiB in either case, as 192 objects or as
6,144. The object count is a choice of representation, not a property of the
data.

\paragraph{Units, callbacks, and paths.}
A \emph{logical cache object} names reusable state and must match the
request's model, representation, tokens, and destination slots; a
\emph{physical byte batch} moves a byte range. Smaller objects refine where a
partial hit ends, while neighboring bytes can still move together. The vLLM
connector exposes the points at which storage can delay or overlap a request:
the start of loading, a per-layer readiness wait, per-layer store callbacks,
and the final save wait~\cite{vllm-connector-contract}. cuFile exposes
host-issued reads and writes on registered handles and buffers; its
compatibility path stages through host memory~\cite{nvidia-cufile-api}. We use
\emph{GDS} for a cuFile transfer whose payload moves between storage and GPU
memory without traversing host DRAM, and judge the path of every run from its
configuration rather than the backend's name~\cite{lmcache-gds-docs}.

\subsection{Where SSD-backed reuse loses time}
\label{sec:bg-loss}

\begin{figure*}[t]
  \centering
  \begin{minipage}[t]{0.46\textwidth}
    \centering
    \includegraphics[width=\linewidth]{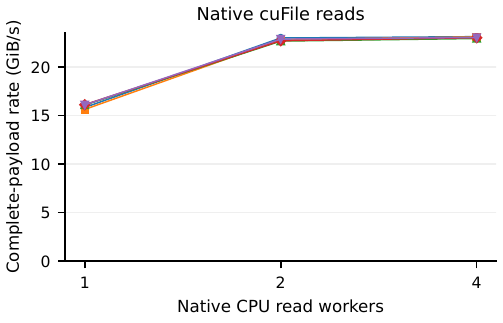}
  \end{minipage}\hfill
  \begin{minipage}[t]{0.50\textwidth}
    \centering
    \includegraphics[width=\linewidth]{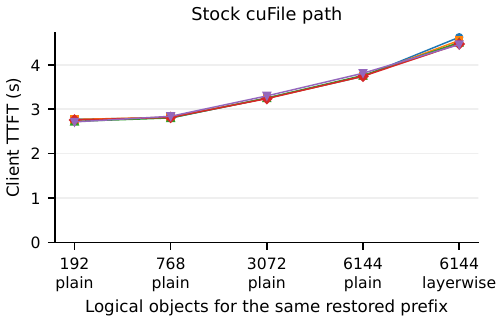}
  \end{minipage}
  \caption{Where the time goes. Left: native host-issued cuFile reads of the
  same 6\,GiB with one, two, and four threads, five interleaved repetitions
  each, every buffer verified; the array ceiling is 22.8\,GiB/s. Right: client
  TTFT of LMCache's cuFile backend restoring the same 6\,GiB prefix as 192,
  768, 3,072, or 6,144 objects and under real layerwise loading, five
  repetitions, all observations shown; read sectors, exact hits, delivered KV,
  and output identity pass for every request.}
  \label{fig:motivation}
\end{figure*}

We restore the same 6\,GiB prefix on the hardware of \S\ref{sec:method} and
vary only how it is partitioned (Figure~\ref{fig:motivation}).

\paragraph{\ding{172} The device is not the bottleneck.}
The native cuFile tool reads the same 6\,GiB as 32\,MiB requests at
\MotNativeOne{}\,GiB/s with one host thread and \MotNativeFour{}\,GiB/s with
four, against a 22.8\,GiB/s array ceiling, with every returned buffer
verified (Figure~\ref{fig:motivation}, left). A host-issued path saturates
the media with a handful of threads; whatever the serving stack loses, it does
not lose it at the drive.

\paragraph{\ding{173} The same bytes cost more as objects multiply.}
Through LMCache's cuFile backend, the same restore takes \MotStockCoarse{}\,s
of client TTFT as \MotStockCoarseObjects{} objects,
\MotStockFine{}\,s as \MotStockFineObjects{} objects, and
\MotStockLayerwise{}\,s under real layerwise loading
(Figure~\ref{fig:motivation}, right). The payload, the model, the cached
tokens, and the GPU compute are identical across the three; only the object
count changes. Layerwise loading, whose purpose is to overlap I/O with
computation, is the slowest of the three because it creates the most objects.

\paragraph{\ding{174} The cost is per object.}
In an offline diagnostic of the same backend, doubling the payload at a fixed
1\,MiB object size (6,144 to 12,288 objects) doubled every line item: the
media phase from 0.325 to 0.655\,s and the serial host phase from 0.418 to
0.819\,s, with a 2\,ms residual. The two phases do not overlap---the host
phase begins when the last read returns---so the host cost lands on the
request's critical path in full, about 67\,\textmu s per object.

\paragraph{\ding{175} Deployments multiply the object count.}
Object count is the product of chunks, layers, and tensor-parallel ranks.
A 128K-token prefix is 512 chunks; on an 80-layer model under layerwise
loading that is 40,960 objects, and with eight-way tensor parallelism, each
rank holding a slice of every object, 327,680. The stock stack charges each of
them at every stage: the batched retrieve API constructs a key, a path, and an
allocation per object before dispatch; each object is a file and a native
call; each object is delivered by its own GPU transfer; the batched store API
iterates over objects and submits each to the scalar path; and each object
adds a record to an index that only grows~\cite{lmcache-gds-source}. A
batch-shaped interface with a per-object loop inside is still per object.

\subsection{Existing answers keep the work per object}
\label{sec:bg-prior}

\paragraph{Move the issuer to the GPU.}
BaM manages NVMe submission and completion queues in GPU memory so that GPU
threads issue fine-grained I/O without CPU involvement~\cite{bam}; GeminiFS
adds host-managed file mappings and a GPU page cache~\cite{geminifs}; CoPilotIO
keeps GPU issuance and moves completion polling to CPU cores, and its uGDS
library exposes CPU-issued userspace NVMe~\cite{copilotio,ugds-source}. Tutti
builds a GPU object store on these primitives, with GPU-issued object I/O and
slack-aware writes, and reports its gains against LMCache's GDS
backend~\cite{tutti}; the LMCache release its text names, 0.4.2, and the 0.4.1
release named in its plots both gate their GDS worker pool on the filesystem
type~\cite{lmcache-gds-old-releases}, and the 0.5.2 release evaluated here
removes that gate. These systems change where an I/O originates. The
per-object work---one command, one completion, one mapping entry per
object---moves with it and remains per object.

\paragraph{Hide the work behind a DRAM stage.}
LMCache's multiprocess tiers serve several engines from one daemon; the
\texttt{fs\_native} adapter issues from a native worker pool, prefetches at
scheduling time, and stages every restored byte through a pinned host slab on
its way to the GPU~\cite{lmcache-mp-source}. HiCache separates host layout
from GPU transfers~\cite{hicache-source}; Mooncake pools KV across
nodes~\cite{mooncake}; py-kvcache combines scheduler lookahead with
event-retired staging~\cite{pykvcache}; Phoenix uses \texttt{ZONE\_DEVICE}
mappings and stream-ordered I/O~\cite{phoenix2025}; \emph{No Buffer, No
Bottleneck} reads CPU-pinned KV through modified attention~\cite{directkv}.
These paths lower or hide the per-object cost---native workers instead of
Python, parallel instead of serial, before the request instead of during
it---and pay for it with host memory bandwidth, pinned capacity, a process
hop, and, in the evaluated multiprocess connector, no layerwise delivery
path at all.

\paragraph{Batch one stage.}
Strata coordinates GPU and host layouts, transfers, indexing, and
scheduling~\cite{strata}; Dual-Blade maps tensors to sequential NVMe
extents~\cite{dualblade}; LMCache's \texttt{raw\_block} backend combines fixed
slots with bounded write groups and aggregated reads~\cite{lmcache-rawblock};
Dynamo KVBM exposes physical layouts with vectorized copies~\cite{kvbm-layout};
ObjectCache separates reuse chunks from layerwise delivery~\cite{objectcache};
SGLANG-LSM separates a prefix-ordered index from tensor logs~\cite{sglanglsm}.
Each batches or packs one stage of the path. A stage that is still per
object---a lookup, a launch, a store submission, an index record---keeps the
whole path per object, because the stages are serial. Reuse policy, which
decides what to compress, retain, or
evict~\cite{cachegen,cacheblend,impress2025,cachedattention,pensieve2025,adaptivekv2026,merlin,lah,three-l-cache,obase,mdk},
is orthogonal: this paper holds reuse decisions and the uncompressed payload
fixed.

\subsection{Challenges of object-count invariance}
\label{sec:bg-challenges}

Making every stage of the path scale with bytes or batches rather than
objects raises three problems that the logical-object representation had
been solving implicitly.

\paragraph{Naming reuse without enumerating objects.}
A partial hit ends at a token boundary that need not align with any batch;
two requests may share a prefix but diverge inside a batch; a batch written by
one request may be read by another before the writer has retired. A batched
path must still answer ``which tokens are reusable, and who owns their
bytes'' exactly, without constructing a key per object (\S\ref{sec:design},
range tickets).

\paragraph{Landing fixed byte batches in a paged, per-layer GPU memory.}
The engine's destination is not a buffer but a set of blocks per layer,
bound to a request only when it is scheduled, and attention for layer $\ell$
must not start before layer $\ell$'s bytes have landed. A fixed-size stripe
on media therefore has to be scattered into slots it did not know at store
time, and released to the model one layer at a time, without turning each
block into its own transfer (\S\ref{sec:design}, byte stripes and delivery).

\paragraph{Owning batches through asynchronous completion.}
One batch spans many logical objects and outlives the call that issued it;
completion is asynchronous on a GPU stream; a request may be cancelled while
its reads are in flight; and a bank reused one step too early silently
corrupts the next request's KV---exactly the failure we observed in a stock
multiprocess GDS pipeline at two requests in flight. Ownership of buffers,
descriptors, and source pages must follow the completion fence, not the API
return (\S\ref{sec:design}, completion-owned banks).

\section{Design and Implementation}
\label{sec:design}

This section presents Range. We first walk one request through the engine
(\S\ref{sec:d-overview}), then describe the three mechanisms that answer the
challenges of \S\ref{sec:bg-challenges}: range tickets, which name reusable
KV without enumerating objects (\S\ref{sec:d-tickets}); physical byte
stripes, which fix the transfer unit independently of reuse granularity
(\S\ref{sec:d-stripes}); and completion-owned transfer banks, which connect
stripes to paged GPU memory and to the model's layer dependencies
(\S\ref{sec:d-banks}). We then give the ownership and recovery rules
(\S\ref{sec:d-ownership}) and the implementation in LMCache and vLLM
(\S\ref{sec:implementation}).

\subsection{Overview}
\label{sec:d-overview}

Range applies one rule to the whole storage path: charge work to payload
bytes and to bounded byte batches, never to logical cache objects. A stripe
layout alone would still construct a key per object; a batched index alone
would still issue object-sized transfers. Range changes both interfaces and
carries the batch representation through lookup, read, delivery, store, and
retirement.

\begin{figure}[t]
 \centering
 \includegraphics[width=\columnwidth]{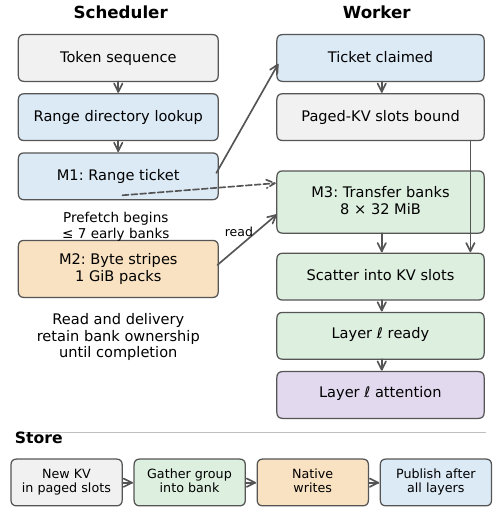}
 \caption{Range's request path. Range tickets carry identity from scheduler
 to worker; physical stripes define I/O; completion-owned banks connect
 storage to paged KV. The dashed edge starts prefetch before destination
 binding. Stores publish after all required layers complete.}
 \label{fig:impl-request-path}
\end{figure}

Figure~\ref{fig:impl-request-path} shows a request with a 49,152-token cached
prefix (6\,GiB) and a 16,128-token suffix on the evaluated model.
\ding{172}~The scheduler sends the packed token sequence to the worker's
range directory, which answers with a \emph{range ticket} naming the reusable
prefix and the six physical packs that hold it; no per-chunk key is built.
\ding{173}~Reads of the first stripes start from this lookup edge, into
up to seven of the eight 32\,MiB read banks, before vLLM has assigned
destination slots. \ding{174}~When the request is scheduled, the worker claims
the ticket and binds token positions to the request's paged-KV slots.
\ding{175}~Four read workers issue 192 stripe reads through cuFile into the
banks; each completed stripe is scattered into its layer's slots by one
transfer, and the bank is released only when the scatter's completion fence
succeeds. \ding{176}~When every stripe of layer $\ell$ has landed, the
coordinator publishes layer $\ell$'s readiness event and attention for that
layer proceeds while later layers are still being read. \ding{177}~As the
suffix is computed, each produced layer group is gathered directly from
paged KV into a 256\,MiB store bank, written to media in 64 native writes for
the 2016\,MiB suffix, and published once all layers have been written. At
every step the operation count is set by bytes and batch geometry: 192 reads,
192 scatters, 32 readiness events, 8 gathers and 64 writes, whether the same
prefix is named as 192 logical objects or 6,144.

\subsection{Range tickets}
\label{sec:d-tickets}

The first challenge of \S\ref{sec:bg-challenges} is to answer ``which tokens
are reusable, and who owns their bytes'' without constructing a key per
object. Range replaces the per-key index with a \emph{range directory}: a
compressed prefix tree over packed token identities whose leaves reference
physical pack records. Lookup takes the scheduler's packed token buffer and a
bounded identity header, compares token bytes against the tree's path labels,
and returns an opaque ticket bound to the request, scheduler session, attempt,
worker rank, namespace, and the storage generations that supply the hit. Index
work follows token-metadata bytes, tree branches, and physical ranges, each
with an explicit capacity bound; for the evaluated unbranched prefix a lookup
selects six pack records, against 192 or 6,144 chunk records in the stock
index (Table~\ref{tab:implementation-work}).

A partial hit ends at a token boundary: the ticket clips the selected physical
ranges at the promised endpoint, so partial-prefix semantics stay an index
decision while transfer aggregation stays a separate one. The connector rounds
the hit to the framework's unchanged logical reuse boundary, so Range cannot
obtain a higher hit rate by changing its layout. Ownership crosses requests
through generations: pack invalidation removes physical records and prunes
dead branches, and a physical lease taken by a ticket survives until delivery
retires, independently of Python garbage collection. The worker claims a
ticket before delivery by checking its envelope, owner epoch, generations,
exact tokens, and promised endpoint; a repeated lookup cannot silently replace
the query behind a ticket. Restoring per-key indexing in the ablation costs
23\,ms at 32\,MiB objects and 911\,ms under layerwise loading
(\S\ref{sec:ablations}).

\subsection{Physical byte stripes}
\label{sec:d-stripes}

The second challenge is to land fixed byte batches in a paged, per-layer GPU
memory. Range stores KV in \emph{packs} inside one preallocated regular file.
A pack holds one stripe per model layer; a stripe is a fixed byte bound $W$
(32\,MiB) of token-major KV in the schema
$[\mathrm{token},\mathrm{K/V},\mathrm{hidden}]$. For pack base $a$, layer
$l$, token offset $t$, and per-layer token size $s$ (4096 bytes for the
evaluated model), the address is $a+lW+ts$; a 32\,MiB stripe holds 8192 tokens
and 32 stripes form a 1\,GiB pack. A value payload follows its token's key
payload, so a partial stripe keeps the same addressing rule and reads or
writes only its valid extent.

\begin{figure}[t]
  \centering
  \includegraphics[width=\columnwidth]{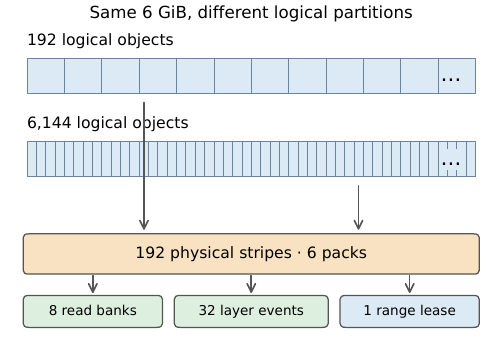}
  \caption{The same 6\,GiB as 192 or 6,144 logical objects maps onto the
  same physical units. The counts are arithmetic from the implementation for
  the recorded unbranched geometry.}
  \label{fig:impl-units}
\end{figure}

Restore and store jobs follow physical spans, not objects. For payload $B$ and
$P$ separately represented partial boundaries, planning and native calls
follow the spans covering $B$ at bound $W$ plus boundary work for $P$;
logical object count does not appear. Every span is checked against file
bounds, registered-buffer bounds, alignment, and storage generation, and
disjoint ranges remain separate spans. Figure~\ref{fig:impl-units} draws the
consequence: the same 6\,GiB, partitioned into 192 or 6,144 logical objects,
maps onto the same 192 stripes, six packs, eight banks, 32 layer events, and
one range lease. Each completed stripe is delivered by one call of the
existing single-layer paged-KV transfer, which scatters the stripe's tokens
into their slots; the stock path invokes the multi-layer transfer once per
object. Replacing coalesced stripe reads with object-sized issuance in the
ablation costs 8\,ms at 32\,MiB objects and 93\,ms under layerwise loading
(\S\ref{sec:ablations}).

\subsection{Completion-owned transfer banks}
\label{sec:d-banks}

The third challenge is to own batches through asynchronous completion.
Bounded read and store banks are carved from the registered GPU allocation at
startup; pool allocation is a one-time cost, and bank admission and
retirement happen per byte batch rather than per object.

\begin{figure}[t]
  \centering
  \includegraphics[width=\columnwidth]{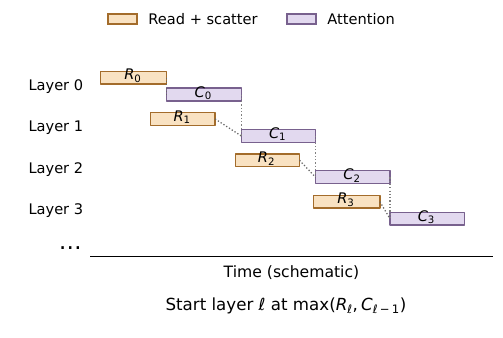}
  \caption{Reading, delivery, and layer consumption. Layer $\ell$ starts at
  $\max(R_\ell, C_{\ell-1})$; banks remain owned through completion. Lane
  lengths are schematic.}
  \label{fig:impl-overlap}
\end{figure}

\paragraph{Reads and layer readiness.}
Delivery uses eight 32\,MiB read banks and four read workers. Lookup prefetch
may occupy seven banks, leaving one for scheduled work; scheduled delivery
looks two layers ahead. Early reads start before destination slots exist, but
a scatter waits for the request's slot binding. The coordinator validates all
layer ranges before issuing any read and checks each synchronous cuFile
return against its requested byte count. Each completed span is scattered on
a request-owned CUDA stream, and the bank remains owned until the scatter's
completion fence succeeds. When every span of a layer has completed, the
coordinator records that layer's readiness; the model stream waits on the
event. Read workers fill other banks while a scatter executes, and later
layers are read while earlier layers compute (Figure~\ref{fig:impl-overlap}).

\paragraph{How the pieces compose.}
Serial dependencies add. If $R_\ell$ is the completion of layer $\ell$'s read
and scatter and $C_{\ell-1}$ the preceding layer's compute, layer $\ell$ starts
at $\max(R_\ell, C_{\ell-1})$. Once a layer's read is hidden behind the
previous layer's compute, advancing it further does not improve TTFT; what
remains on the critical path is the first layer's read and any admission,
extraction, or visibility wait. This is why the measured contribution above a
GPU-resident prefix (\S\ref{sec:evaluation}) is far below the time the media
needs for 6\,GiB, why replacing layer-ready delivery with a whole-prefix
barrier costs about 250\,ms at every point, and why the scheduling-time
prefetch, which advances reads that were already hidden, is worth only a few
milliseconds (\S\ref{sec:ablations}).

\begin{figure}[t]
  \centering
  \includegraphics[width=\columnwidth]{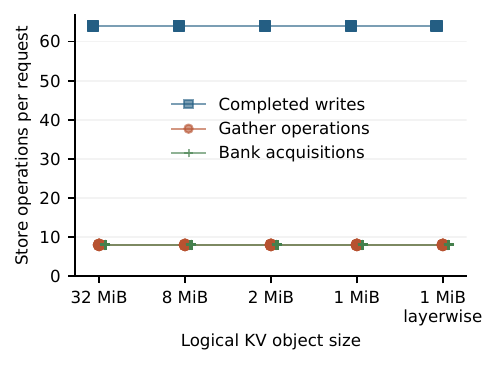}
  \caption[Completed suffix-store work]{\protect\input{figs/final_store_counts.caption}}
  \label{fig:final-store-counts}
\end{figure}

\paragraph{Stores and publication.}
Layer-production callbacks record dependencies on the model stream; a byte
group becomes eligible when its source layers are ready. The store uses eight
256\,MiB banks, eight group workers, and 32 write workers; a full group covers
four layers and at most eight stripes. A vector gather reads the selected
paged GPU slots directly into a registered bank using source pointers,
strides, and token-range descriptors, and cuFile writes the group's physical
spans from that bank; the payload never traverses a host staging pool.
Descriptors and source slots remain owned until extraction completes, banks
until every accepted write returns, and the range becomes visible only after
all required layer writes succeed. Extraction and storage completion are
distinct: extraction releases the model's source slots, successful writes
permit publication, and write debt continues after the connector returns.
Plain and layerwise connector modes share this machinery; the layerwise flag
adds no separate overlap stage. For the 16,128-token suffix (2016\,MiB: one
8192-token pack and a 7936-token tail with 31\,MiB of valid payload per layer),
every request in the sweep completes 64 native writes, eight gathers, and eight
bank acquisitions (Figure~\ref{fig:final-store-counts}).

\subsection{Ownership and recovery}
\label{sec:d-ownership}

Generations, read leases, and completion fences protect physical reuse;
Table~\ref{tab:ownership} lists the owners and the evidence each needs before
it may be retired. Publication requires successful writes for every selected
layer, including groups that complete on different workers; a group failure
prevents visibility, and reclamation waits for all other accepted writes to
drain. Cancellation or an exception is not completion evidence, and an
unproved CUDA fence quarantines its resources rather than releasing them.
Duplicate stores join the existing reservation. Clean shutdown drains owners,
synchronizes payload, and publishes a file-identity-bound metadata snapshot
before clearing the dirty marker; a dirty reopen discards the index rather
than reconstructing it from payload.

\begin{table}[t]
\centering\small
\begin{tabular}{@{}p{0.20\columnwidth}p{0.30\columnwidth}p{0.37\columnwidth}@{}}
\toprule
Owner & State progression & Safe retirement \\
\midrule
Request ticket & lookup $\rightarrow$ claim $\rightarrow$ release & Read futures and scatter fences drained \\
Read bank & free $\rightarrow$ read $\rightarrow$ scatter & Scatter completion proved \\
Store bank & free $\rightarrow$ gather $\rightarrow$ write & All accepted writes returned \\
Pack generation & reserved $\rightarrow$ visible $\rightarrow$ retired & All layers publish; reclaim only at zero refs \\
Dirty marker & open $\rightarrow$ dirty & Remove only after synchronized payload and durable clean snapshot \\
Metadata snapshot & clean $\rightarrow$ reopen & Validate identity/checksum; dirty reopen discards index \\
\bottomrule
\end{tabular}
\caption{Completion ownership in the implementation. An unproved completion retains its owner and prevents reuse. Pack publication and physical reclamation are separate transitions.}
\label{tab:ownership}
\end{table}

\subsection{Implementation}
\label{sec:implementation}

\paragraph{Integration with LMCache and vLLM.}
Range adds 6,889 lines of Python in 34 files under
\texttt{lmcache/v1/range\_engine} and a 71-line CUDA gather kernel to
LMCache's in-process engine. The diff against the LMCache release touches 50
paths ($+8{,}304/-285$ lines); the change to vLLM is four lines in one file
($+3/-1$). The engine source, the patches against LMCache 0.5.2 (commit
\texttt{cd2c0d6a}) and vLLM 0.25.1 (commit \texttt{e04a30a7}), and the
SHA-256 manifest of the installed package (\texttt{77dfae8b\ldots}) are part
of the artifact. The engine lives entirely behind the existing KV-connector
interface and preserves vLLM's block-granular KV management: attention, the
KV dtype, and the paged-cache allocator keep their ordinary implementations.
On the scheduler side, Range extends LMCache's lookup transport; the response
preserves the connector's integer hit API while carrying serialized,
rank-specific tickets, and when vLLM allocates cache slots the adapter moves
the tickets into scheduled-request state and attaches them to the request
sent to the worker. On the worker side, Range implements the connector's
restore callbacks (start of loading and per-layer readiness wait) and store
callbacks (per-layer production and final save wait); the plain and
layerwise connector modes drive the same stripe, bank, and readiness
machinery, differing only in which callbacks the engine invokes. Restores
run on the request's critical path; stores are queued behind
layer-production events and continue after the connector returns. Finishing
or abandoning a request cancels unclaimed lookup ownership and retires its
scheduler state. The coordinator is constructed after LMCache's storage
manager and closes before the registered allocation is disposed.

\paragraph{Identity.}
The namespace binds model revision, semantic configuration, KV dtype,
parallel geometry, and application tags; a change in any of them selects a
separate namespace rather than reusing incompatible persisted tensors.
Scheduler messages carry exact token bytes and request, session, and attempt
identities. After allocation the connector binds an immutable mapping from
token positions to paged-cache slots, so a transfer batch position never
identifies a destination.

\paragraph{Multi-GPU.}
Under tensor parallelism vLLM runs one worker per GPU, and one Range engine
runs in each, with its own pack file, directory, banks, and rank-qualified
namespace. The scheduler's lookup is broadcast to every rank and completes
only when every rank has answered; each rank returns a ticket bound to its own
rank and namespace, and the scheduled request carries the full rank map, so a
worker can claim only the ticket that names it. Each rank's stripes hold that
rank's slice of every layer, so the per-rank operation counts are those of
\S\ref{sec:d-overview} scaled to the slice, and the per-object multiplication
that tensor parallelism imposes on the stock path (\S\ref{sec:bg-loss}) does
not arise. Scheduling-time prefetch is currently enabled only for a single
rank with vLLM's own prefix cache off and no load clipping, because a promise
made at lookup that is later clipped by another rank or by the framework
would need an explicit surplus-read policy. Pipeline parallelism, a local
DRAM tier, cache blending, asynchronous loading, disaggregated transfer, and
per-request LoRA namespaces are rejected at startup rather than silently
degraded. Our evaluation uses one GPU; multi-rank serving is implemented but
not measured in this paper.

\paragraph{Testing.}
The ownership rules of \S\ref{sec:d-ownership} are exercised by CPU fault
tests over temporary pack files that replay stale tickets, divergent
prefixes, partial completion, cancellation during in-flight reads, dirty
reopening, and clean reopening from a snapshot. Every serving request in the
evaluation additionally compares raw-bit fingerprints of all promised K/V
positions across all 32 layers before attention, checks the generated tokens
against a fresh recompute, and verifies physical sector counts and loaded
source hashes.

\paragraph{Work units.}
Table~\ref{tab:implementation-work} lists, from the source of both paths, the
operations a 6\,GiB restore and a 2016\,MiB store perform. Stock's successful
object path allocates a destination and issues a payload read per object, its
batched store calls the scalar submission once per object, and its GPU
connector invokes the native multi-layer transfer once per object; Range
selects pack records and issues one single-layer transfer per physical span,
with temporary payload allocation moved from each object to startup bank
allocation. The counts are source-derived for an unbranched, fully populated
prefix; additional physical fragments add spans.

\begin{table}[t]
\centering\small
\begin{tabular}{@{}lrrr@{}}
\toprule
Source-derived work & Stock & Stock & Range \\
Logical prefix objects & 192 & 6144 & either \\
\midrule
Prefix records selected & 192 & 6,144 & 6 \\
Prefix payload reads & 192 & 6,144 & 192 \\
KV transfer API calls & 192 & 6,144 & 192 \\
Payload allocations & 192 & 6,144 & 0 \\
Suffix write submissions & 63 & 2,016 & 64 \\
\bottomrule
\end{tabular}
\caption{Source-derived successful-path work for the 6\,GiB prefix and actual 2016\,MiB suffix. Stock denotes the plain in-process GDS path. Range selects physical pack records and reuses 8 startup read banks. Transfer API calls span multiple layers in stock and one layer in Range; they do not count internal device launches. Temporary payload allocations exclude metadata and vLLM destination KV.}
\label{tab:implementation-work}
\end{table}

\paragraph{Configuration.}
The installed configuration uses 1\,GiB packs of 32\,MiB stripes, eight read
banks with four read workers and seven prefetch banks, two layers of
lookahead, eight 256\,MiB store banks with eight group workers and 32 write
workers, and a 16\,GiB capacity bound. cuFile settings are shared with the
stock baselines.

\section{Experimental method}
\label{sec:method}

We measure client latency, storage traffic, delivered KV, and process resources
on the same request. A separate loopback HTTP client process times request
submission to the first complete generated-token event. Storage observations
must pass delivered-KV, recompute-output, sector, hit-count, loaded-source,
and process-identity checks. All experiments serve one request at a time; concurrent serving is
future work.

\paragraph{Platform and workload.}
The final experiment uses vLLM~0.25.1 and LMCache~0.5.2 with
Llama-3.1-8B-Instruct and BF16 KV on one H100 NVL. Two NVMe drives form the
ext4 array. The fixed request uses a document excerpt and original question
from L-Eval~\cite{levaldata}. It contains 65,280 tokens, of which 49,152 are cached:
the restore is 6\,GiB and the newly computed suffix is 2,016\,MiB of KV.
Each request asks for two generated tokens; tensor parallelism is one.
The plain partitions contain 192, 768, 3,072, or 6,144 objects; the final
category uses actual layerwise callbacks and 6,144 layer objects. The finest
plain point is a control separating object count from layer dependencies.
Every arm at a point uses the same request and hit boundary.

Recompute and an HBM-resident prefix supply reference paths. The storage arms
are Range, tuned stock in-process cuFile, multiprocess \texttt{fs\_native}
at the coarse points, and in-process O\_DIRECT at the fine points. Stock receives direct
I/O, a tuned worker pool, sufficient staging, and the same disabled cyclic GC
as Range. Shared alignment, lifetime, layerwise-adapter, and integrity repairs
receive no engine credit. The artifact records each allocator and buffer
setting and the exact installed Python and native libraries; changing them
requires a new matched comparison.

\paragraph{Paired observations.}
The evaluation arms were measured on the same day with interleaved order;
the motivation diagnostics were measured earlier on the same machine.
The primary sweep has three attempts per cell, with storage-arm order rotated.
Supplementary MP reruns after a fixture-only repair are marked separately;
they are not interleaved and receive no HBM subtraction.
Each repetition includes a resident-prefix request at the same point.
For arm $a$, point $p$, and repetition $r$, the storage-path contribution is
\begin{equation}
 D_{a,p,r}=T_{a,p,r}-T_{\mathrm{HBM},p,r}.
 \label{eq:storage-contribution}
\end{equation}
It includes changes in overlap, rather than measuring aggregate I/O service.
We take medians after subtraction. Ablations pair each control with the full
engine in two repetitions. Figures show
raw observations and observed ranges. These finite-sample summaries support
neither a distributional model nor confidence intervals. Failures have no
performance value.

\paragraph{Observer placement.}
GPU capture of the delivered KV precedes attention; the CPU comparison and
serialization run after the response for every evaluated arm. Running the
comparison inside the request interval instead inflates TTFT by 0.63\,s for
stock cuFile and 0.66\,s for Range (five repetitions each), so the placement
does not favor either path.

\paragraph{Physical traffic and path identity.}
Warm stores complete before residency removal. Physical read-sector counters
on both array members bracket verification, with the expected payload and
recorded metadata allowance.
Timed processes use cuFile and LMCache ERROR logging, \texttt{PYTHONHASHSEED=0},
CPU affinity 0--31, and captured driver libraries. Only separate,
configuration-matched DEBUG records with native submissions and no fallback
establish GDS. Such records support the native motivation reference; the final
serving sweep is reported as cuFile. Application-copy accounting excludes
unknown driver copies and DRAM-controller traffic; memory capacity is the
configured pool limit, not measured total pinned allocation.

\paragraph{Budget and reproducibility.}
The host budget subtracts paired resident phases from the same request phases
and compares their sum with $D$; the residual is reported in absolute terms.
Overlapping intervals compose through their dependency maximum.
One operator holds the exclusive measurement lock. Imports and startup precede
execution; deadlines, server identity, and process retirement are checked.
Data retain configurations, commands, source hashes, raw values, and failures;
scripts regenerate the figures and ledger.

\section{Serving results}
\label{sec:evaluation}

\begin{figure*}[t]
  \centering
  \includegraphics[width=.98\textwidth]{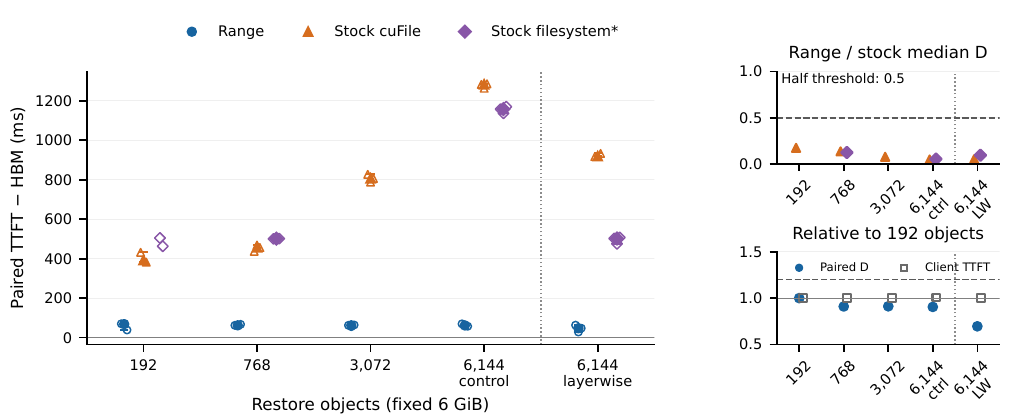}
  \caption{Storage-path contribution relative to resident reuse, measured
  from paired external-client TTFTs in the final sweep. Each observation uses
  its own point and interleaved block's HBM reference. Summary marks show
  medians of those signed differences, with the full observed range. Failed
  cells have no difference value. The control point changes logical object
  count without introducing layer dependencies.}
  \label{fig:primary-contribution}
\end{figure*}

Our evaluation answers three questions. \textbf{Q1:} With the payload,
model, and cached tokens fixed, does TTFT depend on how many logical objects
name the prefix (\S\ref{sec:ev-sweep}--\S\ref{sec:ev-mp})?
\textbf{Q2:} Which mechanism accounts for how much of the result
(\S\ref{sec:ablations})? \textbf{Q3:} What host resources does each path
reserve and move to obtain its latency (\S\ref{sec:resources})?

Range holds median client TTFT between 1.788 and 1.803\,s while logical object count
grows from 192 to 6,144 and delivery becomes layerwise. The spread of point
medians is 15.0\,ms. Tuned in-process cuFile rises from 2.124 to 3.012\,s on
the plain object-count sweep, while Range's finest plain point is only
0.73\% above its coarse point. Figure~\ref{fig:primary-ttft} shows the
complete client result; Figure~\ref{fig:primary-contribution} subtracts the
paired resident-prefix reference.

\subsection{Fixed bytes, increasing object count}
\label{sec:ev-sweep}

\begin{table}[t]
\centering\small
\begin{tabular}{@{}lrrrrr@{}}
\toprule
Objects & 192 & 768 & 3,072 & 6,144 & 6,144 LW \\
Object size & 32\,MiB & 8\,MiB & 2\,MiB & 1\,MiB & 1\,MiB \\
\midrule
Recompute & 4.793 & 4.782 & 4.766 & 4.756 & 4.760 \\
HBM-resident & 1.724 & 1.734 & 1.735 & 1.737 & 1.730 \\
Stock cuFile & 2.124 & 2.184 & 2.538 & 3.012 & 2.664 \\
Stock fs*/O\_DIRECT & 2.206 & 2.235 & 2.320 & 2.895 & 2.235 \\
\textbf{Range} & \textbf{1.790} & \textbf{1.796} & \textbf{1.797} & \textbf{1.803} & \textbf{1.788} \\
\midrule
Range $-$ HBM (ms) & 66 & 62 & 62 & 66 & 58 \\
\bottomrule
\end{tabular}
\caption{Median client TTFT (s) in the fixed-payload sweep (three
repetitions; \texttt{fs\_native} at 192 objects is the median of its two
completed requests and at 3,072 objects the supplementary rerun). Stock
fs* is the multiprocess filesystem path at 192--3,072 objects and O\_DIRECT
at the finer points. The last row is the difference of point medians.}
\label{tab:sweep}
\end{table}

The plain sweep separates logical granularity from the amount of KV restored.
Table~\ref{tab:sweep} lists the point medians. At 192, 768, 3,072, and 6,144 objects, Range's median TTFTs are
1.790, 1.796, 1.797, and 1.803\,s. The corresponding stock cuFile medians
are 2.124, 2.184, 2.538, and 3.012\,s. Recompute remains between
4.756 and 4.793\,s across these points. The payload and request are fixed;
the growing stock line is the serving consequence of its granularity-sensitive
path.

At the finest plain point, Range reduces TTFT by 40.1\% against stock cuFile
and by 37.7\% against the faster stock O\_DIRECT path. The latter comparison
is the stronger denominator at this point: 1.803 versus 2.895\,s.
The non-layerwise fine point is a control that isolates logical object count;
the layerwise point below tests the deployment that naturally creates fine
objects.

\paragraph{Near the resident-prefix reference.}
The paired storage-path contributions for Range are 68.2, 62.0, 62.1, and
61.7\,ms as object count grows. Stock cuFile's contribution increases from
392.4 to 1,283.3\,ms. At 192 objects, Range is within a median 68.2\,ms of
the prefix already being resident in GPU memory. Range schedules prefix
movement alongside suffix execution; the remaining client difference is
small even though the request restores the entire prefix from storage.

The contribution is a latency difference, rather than the duration of the
underlying reads. A small difference is useful precisely because storage
can execute while work common to both requests proceeds. The same-block
subtraction also retains variation in that common work. For this reason the
figure shows every paired observation and computes the median after
subtraction.

\subsection{Layerwise reuse}

Real layerwise Range achieves 1.788\,s TTFT, compared with 2.664\,s for the
stock cuFile path and 2.235\,s for stock direct-disk. Its paired contribution
is 47.4\,ms, the lowest point in the sweep. This reduces TTFT by 20.0\%
against the faster stock path and reduces the paired contribution from
502.0 to 47.4\,ms. Logical layer boundaries therefore preserve the near-resident
result rather than multiplying the cost of the physical storage plan.

The multiprocess filesystem connector has no layerwise delivery path. Its
native workers and early prefetch are useful for whole-prefix loading; they
do not provide the per-layer readiness contract used here. Range applies
the same physical representation to plain and layerwise reuse, with completion
published at the layers that computation can consume.

\subsection{The multiprocess comparison}
\label{sec:ev-mp}

At 768 objects, Range takes 1.796\,s and multiprocess \texttt{fs\_native}
takes 2.235\,s, a 438\,ms reduction. The latter remains a strong stock
architecture: its native pool, queued reads, and scheduler-time prefetch
already avoid exposing much of the repeated work to the request. Its two
completed coarse requests at 192 objects take 2.187 and 2.225\,s. These
observations and the complete 768-object cell show a nearly flat coarse
path, unlike the in-process cuFile line.
After repairing the client fixture, supplementary MP requests at 3,072
objects take a median 2.320\,s (2.318--2.335\,s).

Range also eliminates the application's host-to-GPU restore copy. That
advantage concerns payload movement, rather than all resource costs:
Section~\ref{sec:resources} measures the host memory each path reserves and
the payload each moves through host memory.
The multiprocess design provides a shared tier for serving instances; Range's
evaluation concerns one serving engine using local storage.

The fixed store work in Figure~\ref{fig:final-store-counts} complements these
client measurements: physical work units remain constant as logical objects
multiply. The following ablations measure which controls expose latency on
the request path.

\subsection{Ablations}
\label{sec:ablations}

\begin{figure*}[t]
  \centering
  \includegraphics[width=.98\textwidth]{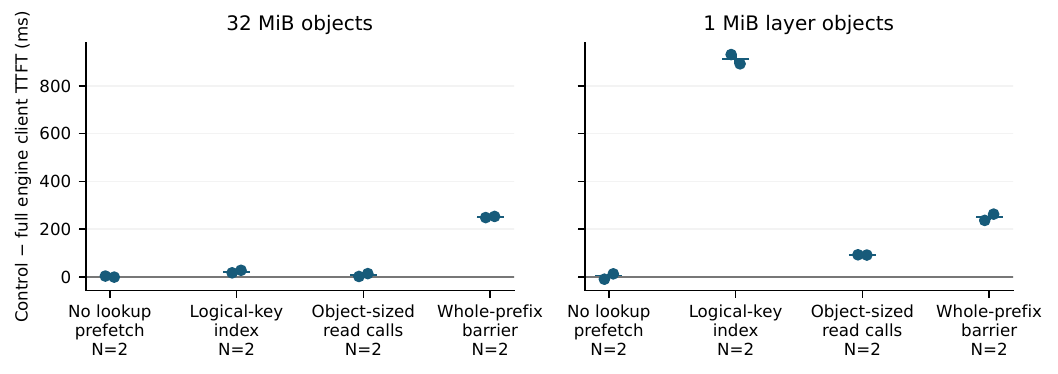}
  \caption{Paired client-TTFT effects of the engine controls. Dots show individual paired differences; bars span the observed range. Positive values mean the control is slower. Each control is labeled with its number of completed pairs. The text describes changes coupled to each control.}
  \label{fig:io-ablations}
\end{figure*}

Figure~\ref{fig:io-ablations} compares each control with the full engine, using 2 repetitions at 2 operating points (20 engine requests including full references). Positive paired differences mean the control is slower. Each pair shares its point, source, request body, and recompute-output reference.

For 32 MiB objects, lookup prefetch disabled changes median TTFT by $+1.861$ ms (range $-0.383$ to $+4.104$ ms); scalar indexing changes median TTFT by $+22.882$ ms (range $+17.582$ to $+28.182$ ms); object-sized read issuance changes median TTFT by $+8.308$ ms (range $+2.341$ to $+14.276$ ms); the whole-prefix barrier changes median TTFT by $+251.038$ ms (range $+248.672$ to $+253.403$ ms).

For 1 MiB layer objects, lookup prefetch disabled changes median TTFT by $+2.101$ ms (range $-9.106$ to $+13.309$ ms); scalar indexing changes median TTFT by $+911.113$ ms (range $+891.718$ to $+930.509$ ms); object-sized read issuance changes median TTFT by $+92.534$ ms (range $+91.970$ to $+93.097$ ms); the whole-prefix barrier changes median TTFT by $+249.937$ ms (range $+236.820$ to $+263.054$ ms).

These controls preserve the physical stripe format, selected KV payload, model, and hit boundary. Object-sized issuance changes native-read partitioning and its planning while preserving each stripe's bank and scatter. Scalar indexing restores key construction, hashing, dictionary access, and key leases together. The prefetch control starts the existing delivery owner at dispatch instead of lookup. The whole-prefix barrier pairs readiness before compute with removal of layer-consumption backpressure, allowing the bounded pool to drain. The byte-group store and extraction/completion contract remain common.

\subsection{Resource cost}
\label{sec:resources}

\begin{figure*}[t]
 \centering
 \includegraphics[width=.7\textwidth]{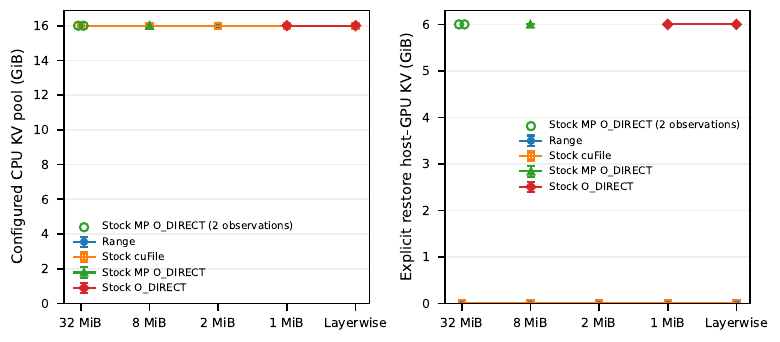}
 \caption{Same-run resource costs. Lines show medians and full ranges of
 three repetitions; open circles show individual observations where an MP
 cell lacks a complete triple. Missing observations reflect client fixture failures.
 Left: configured CPU KV pool capacity; right: source-accounted explicit
 restore payload crossing the host--GPU boundary.}
 \label{fig:resources}
\end{figure*}

Figure~\ref{fig:resources} reports, for the same requests as the sweep,
the host resources each path reserves and moves: the configured CPU KV pool
and the explicit restore payload that crosses the host--GPU boundary.

The primary sweep's storage configurations retain a 16\,GiB CPU KV pool limit.
Source accounting of each measured 6\,GiB restore assigns
6\,GiB of explicit host-to-GPU KV payload to the direct-disk paths and
none to Range or stock cuFile. The application bypasses this restore copy
while retaining the configured CPU pool.

\subsection{An additive host budget}
\label{sec:budget}

The measured host phases explain all but 1.110\,ms of Range's
61.701\,ms contribution at the 1\,MiB point in the actual pair whose
contribution is the median. Their signed sum is 60.591\,ms.
The phases in Figure~\ref{fig:additive-budget} retain the same request's
handoffs: time shifts between model invocation, post-forward work, and
output readiness, so those mixed intervals cannot each be assigned to
storage or compute. Negative changes are retained.

\begin{center}
 \includegraphics[width=.6\columnwidth]{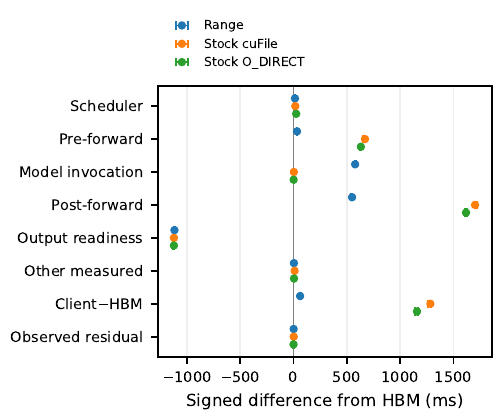}
 \captionof{figure}{Signed host-phase differences from paired HBM at 1\,MiB.
 Each arm shows the phase budget of the pair with median $D$;
 whiskers span the three repetitions; the residual is $D$ minus the measured
 phase sum.}
 \label{fig:additive-budget}
\end{center}

Across the five points and all 15 pairs, the residual is 1--2\,ms for every
path (1.081--1.643\,ms for Range at 1\,MiB), the same order as the end-to-end
client timing resolution; the phase decomposition closes to the millisecond
for Range and stock alike, and the difference in relative terms (2\% of
Range's 60\,ms contribution, 0.1--0.3\% of stock's) comes from the size of
$D$, not from the decomposition.

\paragraph{What limits Range now.}
The 47--68\,ms that separate Range from a GPU-resident prefix are the parts of
the storage path that no batching hides: the first layer's stripe reads, which
cannot overlap any computation because there is no earlier layer to hide
behind; the ticket claim and slot binding, which run once per request before
any read is scattered; and the store's admission and extraction waits, which
gate the first token when the suffix's early layer groups become eligible.
None of these scales with object count, which is why the sweep is flat; they
scale with the number of layers and the size of the first stripe group, which
is where further work on Range would go.

\section{Conclusion}

Range makes fixed-payload KV reuse insensitive to logical object count:
median client TTFT remains 1.79--1.80\,s across a $32\times$ range of object counts
and under layerwise delivery, within 70\,ms of a GPU-resident prefix.
Range tickets, physical byte stripes, and completion-owned transfer banks
carry a byte-or-batch representation through lookup, I/O, GPU delivery,
extraction, and retirement. On the same request, tuned in-process cuFile
rises from 2.124 to 3.012\,s as logical object count grows. The engine
separates reuse granularity from the work required to move a fixed payload,
turning fine-grained KV reuse into a practical serving path.

\raggedbottom
\bibliographystyle{plain}
\bibliography{references,references_frontmatter}

\begin{thebibliography}{10}

\bibitem{obase}
Vinay Banakar, Suli Yang, Kan Wu, Andrea~C. Arpaci-Dusseau, Remzi~H.
  Arpaci-Dusseau, and Kimberly Keeton.
\newblock Obase: Object-based address-space engineering to improve memory
  tiering.
\newblock In {\em OSDI}, 2026.

\bibitem{copilotio}
Guanyi Chen, Qi~Chen, Shu Yin, and Jian Zhang.
\newblock {CoPilotIO}: {CPU} as a co-pilot for {GPU I/O} to free {GPU} compute.
\newblock In {\em 20th USENIX Symposium on Operating Systems Design and
  Implementation (OSDI 26)}, 2026.

\bibitem{impress2025}
Weijian Chen, Shuibing He, Haoyang Qu, Ruidong Zhang, Siling Yang, Ping Chen,
  Yi~Zheng, Baoxing Huai, and Gang Chen.
\newblock {IMPRESS}: An importance-informed multi-tier prefix {KV} storage
  system for large language model inference.
\newblock In {\em 23rd USENIX Conference on File and Storage Technologies},
  FAST '25, pages 187--201. USENIX Association, 2025.

\bibitem{cachedattention}
Bin Gao, Zhuomin He, Puru Sharma, Qingxuan Kang, Djordje Jevdjic, Junbo Deng,
  Xingkun Yang, Zhou Yu, and Pengfei Zuo.
\newblock Cost-efficient large language model serving for multi-turn
  conversations with cachedattention.
\newblock In {\em USENIX ATC}, 2024.

\bibitem{dualblade}
Bodon Jeong, Hongsu Byun, Youngjae Kim, Weikuan Yu, Kyungkeun Lee, Jihoon Yang,
  and Sungyong Park.
\newblock {Dual-Blade}: Dual-path {NVMe}-direct {KV}-cache offloading for edge
  {LLM} inference, 2026.

\bibitem{pykvcache}
Joseph Kanichai, Tiziano~De Matteis, and Animesh Trivedi.
\newblock Building py-kvcache: A performance characterization of external {KV}
  caching for vllm with {NVMe SSDs}, 2026.

\bibitem{merlin}
Liujia Li, Jinhao Guo, Yi~Fan, Jianyu Wu, Zhenlin Wang, Jie Zhang, Yuval Tamir,
  Xiaolin Wang, Yingwei Luo, and Diyu Zhou.
\newblock Merlin: An efficient adaptive cache eviction algorithm via
  fine-grained characterization.
\newblock In {\em OSDI}, 2026.

\bibitem{cachegen}
Yuhan Liu, Hanchen Li, Yihua Cheng, Siddhant Ray, Yuyang Huang, Qizheng Zhang,
  Kuntai Du, Jiayi Yao, Shan Lu, Ganesh Ananthanarayanan, Michael Maire, Henry
  Hoffmann, Ari Holtzman, and Junchen Jiang.
\newblock {CacheGen}: {KV} cache compression and streaming for fast large
  language model serving.
\newblock In {\em ACM SIGCOMM 2024 Conference}, 2024.

\bibitem{lmcache-gds-docs}
{LMCache Project}.
\newblock {GDS Backend}.
\newblock \url{https://docs.lmcache.ai/kv_cache/storage_backends/gds.html}.

\bibitem{lmcache-gds-old-releases}
{LMCache Project}.
\newblock {GdsBackend} in releases 0.4.1 and 0.4.2.
\newblock
  \url{https://github.com/LMCache/LMCache/blob/v0.4.2/lmcache/v1/storage_backend/gds_backend.py}.
\newblock release v0.4.2.

\bibitem{lmcache-gds-source}
{LMCache Project}.
\newblock {GdsBackend} source.
\newblock
  \url{https://github.com/LMCache/LMCache/blob/cd2c0d6a6a982ec5e334bae7704e1029c06d3c97/lmcache/v1/storage_backend/gds_backend.py}.
\newblock commit cd2c0d6a6a98.

\bibitem{lmcache-mp-source}
{LMCache Project}.
\newblock Multiprocess connector and storage adapters.
\newblock
  \url{https://github.com/LMCache/LMCache/blob/cd2c0d6a6a982ec5e334bae7704e1029c06d3c97/lmcache/integration/vllm/lmcache_mp_connector.py#L804-L815},
  2026.
\newblock commit cd2c0d6a6a98.

\bibitem{lmcache-rawblock}
{LMCache Project}.
\newblock Raw block backend and rust data plane.
\newblock
  \url{https://github.com/LMCache/LMCache/tree/c40b7807bd184794492aef4cb8d7d730d5f108b9/rust/raw_block},
  2026.
\newblock commit c40b7807bd18.

\bibitem{directkv}
Shutian Luo and Haiying Shen.
\newblock No buffer, no bottleneck: Efficient zero-copy {KV} cache offloading
  for long-context {LLMs}.
\newblock In {\em 20th USENIX Symposium on Operating Systems Design and
  Implementation (OSDI 26)}, 2026.

\bibitem{nvidia-cufile-api}
{NVIDIA}.
\newblock {GPUDirect Storage cuFile API Reference}.
\newblock
  \url{https://docs.nvidia.com/gpudirect-storage/api-reference-guide/index.html}.
\newblock Sections 3.1--3.3.

\bibitem{kvbm-layout}
{NVIDIA Dynamo Project}.
\newblock {KVBM} physical layouts and transfer implementation.
\newblock
  \url{https://github.com/ai-dynamo/dynamo/tree/c52945c13eba34faa9437326b4377c474e2843f6/lib/kvbm-physical/src/layout},
  2026.
\newblock commit c52945c13eba; the transfer executor and offload pipeline are
  in the same repository.

\bibitem{levaldata}
{OpenLMLab}.
\newblock {L-Eval} document data.
\newblock
  \url{https://github.com/OpenLMLab/LEval/tree/cd34b050269148aed75acbbe4a599873ad0f37e9}.
\newblock commit cd34b0502691.

\bibitem{mdk}
Shaurya Patel, Suli Yang, Yawen Wang, Kan Wu, Alexandra~(Sasha) Fedorova, Margo
  Seltzer, and Kimberly Keeton.
\newblock Mdk: Rethinking the data center memory reclamation problem.
\newblock In {\em OSDI}, 2026.

\bibitem{mooncake}
Ruoyu Qin, Zheming Li, Weiran He, Jialei Cui, Feng Ren, Mingxing Zhang, Yongwei
  Wu, Weimin Zheng, and Xinran Xu.
\newblock {Mooncake}: Trading more storage for less computation---a
  {KVCache}-centric architecture for serving {LLM} chatbot.
\newblock In {\em 23rd USENIX Conference on File and Storage Technologies (FAST
  25)}, 2025.

\bibitem{tutti}
Shi Qiu, Yifan Hu, Xintao Wang, Wenhao Zhu, Jianqin Yan, Hao Chen, Kaiqiang Xu,
  Kai Chen, and Yiming Zhang.
\newblock {Tutti}: Making {SSD}-backed {KV} cache practical for long-context
  {LLM} serving.
\newblock \url{https://arxiv.org/abs/2605.03375v1}, 2026.
\newblock Version 1.

\bibitem{geminifs}
Shi Qiu, Weinan Liu, Yifan Hu, Jianqin Yan, Zhirong Shen, Xin Yao, Renhai Chen,
  Gong Zhang, and Yiming Zhang.
\newblock {GeminiFS}: A companion file system for {GPUs}.
\newblock In {\em 23rd USENIX Conference on File and Storage Technologies (FAST
  25)}, 2025.

\bibitem{bam}
Zaid Qureshi, Vikram~Sharma Mailthody, Isaac Gelado, Seungwon Min, Amna Masood,
  Jeongmin Park, Jinjun Xiong, C.~J. Newburn, Dmitri Vainbrand, I-Hsin Chung,
  Michael Garland, William Dally, and Wen mei Hwu.
\newblock {GPU}-initiated on-demand high-throughput storage access in the {BaM}
  system architecture.
\newblock In {\em Proceedings of the 28th ACM International Conference on
  Architectural Support for Programming Languages and Operating Systems, Volume
  2}, 2023.

\bibitem{ugds-source}
{ScaleX-IO}.
\newblock {uGDS}: User-space {GPU} direct storage development kit.
\newblock
  \url{https://github.com/ScaleX-IO/uGDS/tree/79643ecc3e1841cff289eacbd522cd5c812da313},
  2026.
\newblock commit 79643ecc3e18.

\bibitem{hicache-source}
{SGLang Project}.
\newblock {HiCache}: Host layout and cache controller.
\newblock
  \url{https://github.com/sgl-project/sglang/tree/03ea13a54557de52da5faab2c422da07c3727407/python/sglang/srt/mem_cache},
  2026.
\newblock commit 03ea13a54557.

\bibitem{adaptivekv2026}
Yang Shen, Meghana Madhyastha, Robert Underwood, Bogdan Nicolae, and Randal
  Burns.
\newblock Recency/frequency adaptive {KV} caching for large language model
  serving.
\newblock In {\em Resource-Adaptive Foundation Model Inference Workshop at
  ICML}, AdaptFM '26, 2026.

\bibitem{vllm-connector-contract}
{vLLM Project}.
\newblock {KVConnectorBase} interface.
\newblock
  \url{https://github.com/vllm-project/vllm/blob/e04a30a77cb17d5a9753c53c51ee30bba887053c/vllm/distributed/kv_transfer/kv_connector/v1/base.py}.
\newblock commit e04a30a77cb1.

\bibitem{lah}
Haocheng Xia, William Nixon, Bintang~Dwi Marthen, Pranav Bhandari, and Juncheng
  Yang.
\newblock Learning-augmented heuristics: Simple, yet smart, robust and
  interpretable cache eviction.
\newblock In {\em OSDI}, 2026.

\bibitem{strata}
Zhiqiang Xie, Ziyi Xu, Mark Zhao, Yuwei An, Vikram~Sharma Mailthody, Scott
  Mahlke, Michael Garland, and Christos Kozyrakis.
\newblock {Strata}: Hierarchical context caching for long context language
  model serving.
\newblock In {\em 20th USENIX Symposium on Operating Systems Design and
  Implementation (OSDI 26)}, 2026.

\bibitem{phoenix2025}
Jianqin Yan, Shi Qiu, Yina Lv, Yifan Hu, Hao Chen, Zhirong Shen, Xin Yao,
  Renhai Chen, Jiwu Shu, Gong Zhang, and Yiming Zhang.
\newblock {Phoenix}: A refactored {I/O} stack for {GPU} direct storage without
  phony buffers.
\newblock In {\em Proceedings of the International Conference for High
  Performance Computing, Networking, Storage and Analysis}, 2025.

\bibitem{cacheblend}
Jiayi Yao, Hanchen Li, Yuhan Liu, Siddhant Ray, Yihua Cheng, Qizheng Zhang,
  Kuntai Du, Shan Lu, and Junchen Jiang.
\newblock {CacheBlend}: Fast large language model serving for {RAG} with cached
  knowledge fusion.
\newblock In {\em Twentieth European Conference on Computer Systems (EuroSys
  25)}, 2025.

\bibitem{pensieve2025}
Lingfan Yu, Jinkun Lin, and Jinyang Li.
\newblock Stateful large language model serving with {Pensieve}.
\newblock In {\em Proceedings of the Twentieth European Conference on Computer
  Systems}, EuroSys '25. Association for Computing Machinery, 2025.

\bibitem{sglanglsm}
Weiping Yu, Ye~Jiarui, He~Mengke, Junfeng Liu, and Siqiang Luo.
\newblock On 10x better scalability: {KV} stores scale up {KV} cache, 2025.

\bibitem{three-l-cache}
Wenbin Zhou, Zhixiong Niu, Yongqiang Xiong, Juan Fang, and Qian Wang.
\newblock 3l-cache: Low overhead and precise learning-based eviction policy for
  caches.
\newblock In {\em FAST}, 2025.

\bibitem{objectcache}
Yu~Zhu, Aditya Dhakal, Yunming Xiao, Dejan Milojicic, and Gustavo Alonso.
\newblock {ObjectCache}: Layerwise object-storage retrieval for {KV} cache
  reuse, 2026.

\end{thebibliography}
\end{document}